\documentclass[%
 aip,
 amsmath,amssymb,
reprint,%
floatfix,
]{revtex4-1}

\usepackage{graphicx}
\usepackage{dcolumn}
\usepackage{bm}
\usepackage{longtable}

\usepackage[utf8]{inputenc}
\usepackage[T1]{fontenc}
\usepackage{mathptmx}
\usepackage{etoolbox}
\usepackage{siunitx, upgreek}
\usepackage[version=4,arrows=pgf-filled,
mathfontname=mathsf]{mhchem}
\DeclareMathAlphabet{\mathcal}{OMS}{cmsy}{m}{n}
\SetMathAlphabet{\mathcal}{bold}{OMS}{cmsy}{b}{n}

\makeatletter
\def\@email#1#2{%
 \endgroup
 \patchcmd{\titleblock@produce}
  {\frontmatter@RRAPformat}
  {\frontmatter@RRAPformat{\produce@RRAP{*#1\href{mailto:#2}{#2}}}\frontmatter@RRAPformat}
  {}{}
}%
\makeatother
\begin{document}

\preprint{AIP/123-QED}

\title[Inferring Thermal Dissociation Kinetics of Ion Clusters from Molecular Dynamics]{Inferring Thermal Dissociation Kinetics of Ion Clusters from Molecular Dynamics}

\author{Nicholas Laws}
 \email{nrl49@cornell.edu}
\author{Adler Smith}%

\author{Elaine Petro}
\affiliation{ 
Sibley School of Mechanical \& Aerospace Engineering, Cornell University, Ithaca New York
}%

\date{\today}

\begin{abstract}
    Electrospray ionic liquid ion sources operating in the pure ion regime emit metastable, low solvation number ion clusters whose post-emission dissociation modifies plume composition, energy distributions, and diagnostic observables. While field-free measurements support first-order, thermally activated decay on microsecond timescales, the acceleration region is characterized by rapidly varying electric fields and nanosecond residence times such that dissociation becomes strongly field-enhanced. In this work, microcanonical ensemble molecular dynamics (MD) trajectories are used to infer dissociation kinetics of \ce{EMI-BF4} ion clusters under controlled internal energy and applied electric field. Positive and negative dimers, trimers, and tetramers are simulated at temperatures between $600$ to $1000$ K and uniform fields of $10^6$ to $10^9$~\unit{\volt/\m}. Dissociation lifetimes are extracted using a connectivity-based fragmentation criterion, and product channels are classified to obtain pathway-resolved branching probabilities. Across all solvation numbers, lifetimes are field independent in the weak field limit, but collapse by orders of magnitude above $10^8$~\unit{\volt/\m}. Cluster polarity dependence is most prominent in the temperature-controlled regime and diminish as electrostatic work dominates. The MD simulations further reveal that field-driven changes in dissociation topology, including a transition in trimer breakup from neutral pair evaporation to charged core ion ejection, and three competing tetramer pathways involving single neutral loss, double neutral emission, and core ion ejection. The MD dataset is reduced to compact parameterizations of $\tau_n\left(E,T\right)$, yielding transferable kinetic inputs for multiscale plume transport models.
\end{abstract}

\maketitle

\section*{Nomenclature}
\vspace{-.5em}
{
\renewcommand{\arraystretch}{1.2}%

\begin{longtable}{rl{c}}

$n$           & = Cluster solvation number \\
$N_n$           & = Number of survival ion clusters \\
$k_n^0$           & = Field-free first-order dissociation rate constant \\
$S_n$           & = Survival probability \\
$\tau_n^0$           & = Temperature and field dependent cluster mean lifetime\\
$\Delta G_n^{\ddagger}$           & = Effective activation free energy \\
$k_B$           & = Boltzmann constant \\
$A_n$           & = Arrhenius prefactor \\
$\Delta W_n$           & = Field-induced barrier lowering \\
$T$ & = Temperature \\
$E$           & = Electric field \\
$N$ & = Number of independent molecular dynamics trajectories\\
$U_{\mathrm{tot}}$ & = Total potential energy \\
$r$ & = Bond length \\
$r_0$ & = Equilibrium bond length \\
$k_r$ & = Bond force parameter \\
$\theta$ & = Bond angle \\
$\theta_0$ & = Equilibrium bond angle \\
$k_\theta$ & = Angle force parameter \\
$\phi$ & = Dihedral angle \\
$k_\phi$ & = Dihedral force parameter \\
$n_\phi$ & = Dihedral multiplicity \\
$\lambda$ & = Dihedral phase factor\\
$\psi$ & = Improper dihedral angle\\
$\psi_0$ & = Equilibrium improper dihedral angle\\
$r_{ij}$ & = Distance between atoms $i$ and $j$\\
$\varepsilon_{ij}$ & = Lennard-Jones well depth for atoms $i$ and $j$\\
$\sigma_{ij}$ & = Lennard-Jones size for atoms $i$ and $j$\\
$\varepsilon_{i}$ & = Lennard-Jones well depth for atom $i$\\
$\sigma_{i}$ & = Lennard-Jones size for atom $i$\\
$q_{i}$ & = Partial charge for atom $i$\\
$\varepsilon_{0}$ & = Vacuum permittivity\\
$d_b$ & = Distance threshold defining bond adjacency\\
$k_{frag}$ & = NVE trajectory frame index where fragmentation is detected\\
$d_f$ & = Intermolecular separation threshold\\
$q$ & = Cluster charge \\
$G_{IPM}\left(E\right)$ & = Image potential barrier lowering work \\
$G_{DPM}\left(E\right)$ & = Dipole potential barrier lowering work \\
$x$ & = Equilibrium center-of-mass separation\\
$d$ & = Effective dipole length \\
$E_c$ & = Crossover electric field between IPM and DPM \\
$\alpha\left(E\right)$ & = Mixing weight to interpolate between IPM and DPM regimes \\
$G\left(E\right)$ & = Field-weighted barrier lowering work \\
$\Delta G_n^{0}$           & = Zero-field dissociation energy \\
\end{longtable}
}

\section*{Acronyms}
\vspace{-.5em}
{
\renewcommand{\arraystretch}{1.2}%

\begin{longtable}{rl{c}}

ILIS           & = Ionic liquid ion source\\
SSE           & = Secondary species emission \\
MD           & = Molecular dynamics \\
ILIS           & = Ionic liquid\\
PIR           & = Pure ion regime \\
EP           & = Electric propulsion \\
TCID           & = Threshold collision-induced dissociation \\
BDE           & = Bond dissociation energy \\
BDE           & = Particle-in-cell \\
RPA & = Retarding potential analyzer \\
LJ           & = Lennard-Jones \\
IPM           & = Image potential model \\
DPM           & = Dipole potential model \\

\end{longtable}
}

\setcounter{table}{0}
\section{\label{sec:level1}Introduction}
\subsection{Motivation}

Electrospray ionic liquid ion sources (ILIS) enable compact, high specific impulse propulsion for miniature spacecraft and provide a chemically versatile alternative to conventional ion sources for precise ion beam applications \cite{lozano2005ionic, zorzos2008use}.

The central challenge shared across these applications is that the emitted beam does not remain compositionally static after extraction. In the pure ion operating regime, ILIS emits a mixture of low solvation number molecular ions whose subsequent dissociation produces additional charged and neutral products during acceleration. Intra-plume processes and unimolecular dissociation of ions alter the beam’s species content and energy distribution, increases beam divergence that prompts secondary species emission (SSE), and fundamentally couples source performance to the kinetics of the ion cluster dissociation \cite{petro_2022, schroeder2023inferring}. 

Downstream of the emission source, field-free measurements indicate that solvated ion clusters dissociate approximately as an activated, unimolecular process with mean lifetimes on the order of microseconds \cite{miller2020measurement}. Although, the regions in which beam energy is established are characterized by rapidly varying electric fields and short residence times such that effective dissociation kinetics are strongly field-assisted and difficult to isolate experimentally. Recent diagnostics studies have begun to constrain dissociation behavior within the acceleration environment, yet the inversion of measured energy and composition remain model dependent \cite{lyne2024inferring}. Therefore, multiscale simulation tools for electrospray plumes and diagnostics must treat dissociation kinetics as an external input. Multiscale plume models have demonstrated that predicted energy distributions and diagnostic observables can be highly sensitive to assumed field and energy dependence of unimolecular fragmentation probability \cite{nuwal2021multiscale, petro_2022, schroeder2023inferring}. This limitation inhibits extrapolation across operating voltage, geometry, and propellant chemistry, and complicates the development of transferable models that can be validated across experiments.

Molecular dynamics (MD) provides a complementary route to address this gap by directly sampling the atomistic dynamics of solvated ion clusters under controlled internal energy and applied field conditions. Prior MD studies have established the feasibility of inferring dissociation behavior and thermal stability of ionic liquid (IL) clusters and have motivated transition-state interpretations of unimolecular dissociation pathways \cite{iribarne1976evaporation, prince2015molecular, schroeder2023inferring, tahsin2025inferring}. The present work provides a framework for inferring thermal, field-assisted dissociation of ion clusters by leveraging MD. Through the development of a systematic methodology to extract field and temperature dependent dissociation rates and pathways branching from large ensembles of trajectories, we present reduced-form parameterizations of ion cluster mean lifetimes for direct integration into higher dimensional electrospray plume models. By producing pathway-resolved kinetics, this work supplies the transferable atomistic inputs needed to improve model accuracy across all electric field operating regimes.

\subsection{Background}
\subsubsection{Pure Ion Operating Regime} \label{sec:pure_ion_regime}
Electrospray ILIS have been demonstrated across multiple emitter architectures and propellants to generate beams composed predominantly of low solvation number molecular ions under nominal operating conditions \cite{lozano2005ionic, gassend2009microfabricated, matsukawa2023emission, dworski2024investigating, villegas2024emission}. This operating mode is commonly termed the pure ion regime (PIR) and is distinguished from mixed and droplet-emitting modes by the dominance of discrete ionic species rather than charged droplets. PIR operation is of interest for electric propulsion (EP) applications because the highest charge-to-mass ratio clusters maximize specific impulse and propulsive efficiency, while also producing beam conditions amenable for precise experimental diagnostics. For ILs such as \ce{EMI-BF4}, PIR emission consists primarily of \ce{EMI+} or \ce{BF4-} monomers and small solvated clusters produced through ion evaporation from the apex of an electrically stressed meniscus, where the electric field is geometrically concentrated near the emission site \cite{lozano2005ionic, coffman2016electrically, krejci2017emission, gallud2022emission}. Positive and negative polarity operations are obtained by reversing the applied potential between the emitter and a downstream extraction electrode, which enables bipolar ion emission where the dominant emitted cluster families are respectively cations or anions \cite{lozano2005ionic}. 

Electrospray sources operating in the PIR generate predominantly singly charged solvated ion clusters. These clusters are categorized using a solvation number nomenclature that counts the number of neutral ion pair units bound to a charge core ion. A single bare ion with no bound neutral pair is classified as a monomer such that $n=0$, while a charge core bound to one neutral pair is termed a dimer such that $n=1$. Larger clusters follow the same pattern where trimers are denoted as $n=2$, tetramers are prescribed as $n=3$, and so on. For \ce{EMI-BF4}, the corresponding singly charged cluster series may be written as \ce{[EMI-BF4]_{n} EMI+} in positive polarity and \ce{[EMI-BF4]_{n} BF4-} in negative polarity. Representative PIR measurements for \ce{EMI-BF4} regularly indicate that monomers and dimers comprise the dominant beam populations with trimers contributing a smaller fraction \cite{lozano2005ionic, villegas2024emission, ulibarri2025direct}. Although, the precise distribution of species in the electrospray plume varies with operating voltage, feed conditions, emitter geometry, and propellant properties \cite{krejci2017emission, petro2020characterization, petro_2022}. Higher mass ion clusters are also present at low but non-negligible percentages \cite{jia2022quantification, ulibarri2025direct}. The quantitative inference of higher mass ion clusters from measurements remains experimentally challenging and is an active topic of research. Table~\ref{tab:cluster_species} summarizes the nomenclature, composition, and molar masses of singly charged \ce{EMI-BF4} clusters in both polarities. The emitted species provide the discrete molecular clusters whose post-emission dissociation kinetics are inferred from MD trajectories in this work. 

\begin{table}[ht]
    \caption{\label{tab:cluster_species}
    Solvation number, composition, and molar mass of singly charged \ce{EMI-BF4} ion clusters.
    Positive polarity species are listed on the first line of each cell and negative polarity species on the second.}
    \begin{ruledtabular}
    \renewcommand{\arraystretch}{1.15}
    \begin{tabular}{lccc}
    Solvation & Cluster & Composition & Mass \\
    Number & Species & & [\unit{\gram/\mol}]\\
    \hline \\[-1.75ex]
    0 & Monomer  & \shortstack{\ce{EMI+}\\\ce{BF4-}} &
                   \shortstack{$111$\\$87$} \\
    1 & Dimer    & \shortstack{\ce{[EMI-BF4]EMI+}\\\ce{[EMI-BF4]BF4-}} &
                   \shortstack{$309$\\$285$} \\
    2 & Trimer   & \shortstack{\ce{[EMI-BF4]2EMI+}\\\ce{[EMI-BF4]2BF4-}} &
                   \shortstack{$507$\\$483$} \\
    3 & Tetramer & \shortstack{\ce{[EMI-BF4]3EMI+}\\\ce{[EMI-BF4]3BF4-}} &
                   \shortstack{$705$\\$681$} \\
    4 & Pentamer & \shortstack{\ce{[EMI-BF4]4EMI+}\\\ce{[EMI-BF4]4BF4-}} &
                   \shortstack{$903$\\$879$} \\
    \end{tabular}
    \end{ruledtabular}
\end{table}

Once ions are evaporated from the liquid meniscus interface, they are accelerated downstream by the electric field established between the emitter and extractor electrode. Continuum electrohydrodynamic simulations of emitting IL menisci indicate that the local electric field at the emission site for PIR ion evaporation can reach values on the order of $10^{9}$~\unit{\volt/\meter} and then rapidly decays with distance due to extractor boundary conditions \cite{coffman2016electrically, gallud2022emission}. Although, a significant field persists downstream because ions must transit an extractor aperture and the electrostatic potential relaxes over a finite distance, leaving field magnitudes on the order of $10^{6}$ to $10^{7}$~\unit{\volt/\meter} in portions of the near plume for ILIS geometries \cite{nuwal2021multiscale, huh2022simulation, takagi2024simple}. In the present work, the region where ions experience a significant fraction of the potential drop is termed the acceleration region, while the downstream region where the electric field is sufficiently small that additional acceleration is negligible is termed the field-free region. 

\subsubsection{Thermal Unimolecular Dissociation of Solvated Ion Clusters}

Solvated ion clusters emitted in the PIR are metastable molecular aggregates whose composition can evolve during transport. For the singly charged cluster families introduced in Section~\ref{sec:pure_ion_regime}, the dominant dissociation mechanism is described as unimolecular evaporation of a neutral cation-anion pair, which decreases the solvation number by one. In positive polarity, this process can be written schematically as: \ce{[EMI-BF4]}$_{n}$\ce{EMI+}\ce{->[k_{n}]}\ce{[EMI-BF4]}$_{n-1}$\ce{EMI+}\ce{+}\ce{EMI-BF4}, with an analogous expression for negative polarity clusters. Because the products remain molecular ions, sequentially dissociation events are possible for larger parent clusters ($n \ge 2$), leading to cascades of decreasing solvation number as the plume diverges and accelerates. The rate at which these unimolecular events occur is a primary reasoning of the downstream species mixture and the resulting velocity distributions. The present work is motivated by the need to describe cluster stability in terms of a physically meaningful lifetime \cite{hogan2010ion, prince2015molecular, miller2020measurement}.  

In the field-free region, where the electric field is sufficiently weak that additional polarization work on the cluster is negligible, dissociation is commonly modeled as a first-order decay process with a constant rate for a given cluster type and internal temperature. Equation~\ref{eq:survival_fraction} describes the empirical first-order dissociation process for solvated ion clusters.

\begin{equation}
    \begin{aligned}
    \frac{dN_n}{dt} = -k_n^0(T)\,N_n
    \end{aligned}
    \label{eq:survival_rate}
\end{equation}

\begin{equation}
    \begin{aligned}
    S_n(t) = \frac{N_n(t)}{N_n(0)} = \exp\!\left[-k_n^0(T)\,t\right]
    \end{aligned}
    \label{eq:survival_fraction}
\end{equation}

\begin{equation}
    \begin{aligned}
    \tau_n^0(T)=\frac{1}{k_n^0(T)}
    \end{aligned}
    \label{eq:free_lifetime}
\end{equation}

where $N_n\left(t\right)$ indicates the number of survival ion clusters of solvation number $n$, $k_n^0$ represents the field-free unimolecular dissociation rate constant, $S_n\left(t\right)$ denotes the survival probability, and $\tau_n^0$ is the mean lifetime. The first-order form is strongly supported by direct measurements in ILIS beams downstream of the extractor, where cluster dissociation occurs on microsecond timescales and is well described by constant-rate kinetics over relevant flight distances \cite{miller2020measurement}. 

The strong temperature sensitivity of $\tau_n^0$ motivates the use of the Eyring-Polanyi kinetic theory as described in Equation~\ref{eq:eyring}

\begin{equation}
    \begin{aligned}
    k_n^0(T) = A_n(T)\,\exp\!\left(-\frac{\Delta G_n^{\ddagger}}{k_B T}\right),
    \end{aligned}
    \label{eq:eyring}
\end{equation}

where $\Delta G_n^{\ddagger}$ is the effective activation free energy, $k_B$ is the Boltzmann constant, and $A_n\left(T\right)$ is the Arrhenius prefactor. Therefore, field-free rate measurements provide an avenue to infer effective dissociation barrier and post-emission internal temperatures via the inversion of Equation~\ref{eq:eyring}, provided that the cluster remains in an approximately field-free environment over the measurement region \cite{miller2020measurement}. Several complementary experimental and computational studies corroborate this thermally activated, unimolecular mechanism. Miller et al. directly measured dissociation rates of solvated ions for multiple ILs and reported microsecond lifetimes in the downstream, weak-field region with rates increasing rapidly with temperature \cite{miller2020measurement}. Hogan et al. performed mass spectrometry studies of isolated IL clusters, observing that sequential neutral pair evaporation is the primary dissociation mechanism in the field-free regime and that ion cluster stability decreases with increasing solvation number \cite{hogan2010ion}. Gas-phase cluster energetic studies using threshold collision-induced dissociation (TCID) further constrain bond dissociation energies (BDE) and provide atomistic evidence for the binding strengths and dissociation energetics \cite{roy2020gas}. These results support the use of field-free, first-order survival statistics and activated temperature dependencies as an empirically grounded baseline for ILIS cluster dissociation.

In the acceleration region, emitted ion clusters experience significant, spatially varying electric fields while their translational kinetic energy is being established. In this region, dissociation is field-assisted, meaning the external field polarizes the cluster and performs electrostatic work along the dissociation coordinate, lowering the effective barrier for evaporation. Classical ion evaporation of charge droplets in strong field emission, where the free-energy barrier is reduced by a Schottky electrostatic contribution depends on the field magnitude \cite{iribarne1976evaporation, loscertales1995experiments} as described in Equation~\ref{eq:field_assisted}

\begin{equation}
    \begin{aligned}
    k_n(E,T) = A_n(T)\,\exp\!\left[-\frac{\Delta G_n^{\ddagger}(E)}{k_B T}\right]
    \end{aligned}
    \label{eq:field_assisted}
\end{equation}

\begin{equation}
    \begin{aligned}
   \Delta G_n^{\ddagger}(E)=\Delta G_n^{\ddagger}(0)-\Delta W_n(E)
    \end{aligned}
    \label{eq:effective_barrier_field_assisted}
\end{equation}

where $\Delta W_n\left(E\right)$ denotes the field-induced barrier lowering. Equation~\ref{eq:field_assisted} emphasizes that dissociation kinetics in the acceleration region depend jointly on the internal temperature, local electric field magnitude, and cluster solvation number. Field-assisted dissociation is challenging to experimentally measure because residence times through the high field regime of the acceleration region can be on the order of nanoseconds, while the field strength varies by order of magnitude over the comparable distances. 

Experimental diagnostics that infer dissociation behavior within or near the acceleration region often require computational modeling that infers kinetics without additional constraints \cite{petro_2022, schroeder2023inferring, lyne2024inferring}. Recent work has made substantial progress in constraining acceleration region dissociation kinetics, while also emphasizing the sensitivity of observable metrics to the assumed functional form of $k_n\left(E,T\right)$. Nuwal et al. coupled MD simulations of \ce{EMI-BF4} ion cluster stability with 3D particle-in-cell (PIC) plume simulations \cite{nuwal2021multiscale}. Nuwal demonstrates that simulated retarding potential analyzer (RPA) curves are highly sensitive to fragmentation probability, indicating there is strong incentive to characterize $k_n\left(E,T\right)$. Their MD analysis further informs that, for \ce{EMI-BF4} dimers in the $300$ to $600$ K range, fragmentation is suppressed below a threshold field and becomes significant only as the field increases toward the \unit{\volt/\nano\meter} scale. This suggests that thermal activation and field-assisted barrier lowering jointly control dissociation in the near plume region \cite{nuwal2021multiscale}. Petro et al. introduced an electrospray emission framework that couples an EHD meniscus model to a gridless $N$-body plume simulation toolkit with metastable cluster transport \cite{petro_2022}. In this approach, dissociation is treated explicitly during near-field acceleration, and comparison of simulated and measured retarding potential characteristics enables inference of internal cluster temperatures and dissociation behavior. Building on this work, Schroeder et al. developed a coordinated MD and simulated RPA analysis workflow to infer electrospray emission characteristics from measured energy distributions \cite{schroeder2023inferring}. Schroeder identifies a crossover from a temperature-controlled regime, where internal energy dominates and field effects are weak, to a field-controlled regime where electrostatic work dominates and field-assisted barrier lowering strongly modifies lifetimes. Most recently, Lyne et al. reported species-resolved constraints on acceleration region lifetime using energy-resolved mass spectrometry of an \ce{EMI-BF4} plume and an analytical rate model to invert measured ion energy distributions \cite{lyne2024inferring}. By directly linking the energy dispersion of specific cluster species to their probability of dissociation during acceleration, this approach provides a direct window into nanosecond timescale lifetime and yields kinetic parameters that complement RPA inversion methods. Taken together, these studies establish that ILIS cluster dissociation is well described as thermally activated and approximately first order in weak fields, but becomes strongly field-assisted during acceleration. Neutral products generated by prompt cluster dissociation can escape conventional experimental diagnostics, potentially contributing to the discrepancy between directly measured and inferred propellant mass flow \cite{naemura2025direct}. Quantifying this contribution requires field and temperature-dependent dissociation lifetimes and pathway branching fractions to predict when neutral products form and the amount of mass they carry. In this work a robust, transferable procedure for extracting pathway-resolved $k_n\left(E,T\right)$ from atomistic simulations is presented to be utilized in higher dimension plume models.

\section{Methodology}

\subsection{Computational Model of Electrospray Ion Cluster Thermal Dissociation Kinetics}
\subsubsection{Simulation Domain for Ion Cluster Dissociation Kinetics}

Classical molecular dynamics (MD) simulations are performed to resolve the atomistic breakup of solvated \ce{EMI-BF4} ion clusters under combined thermal loading and externally applied electric fields. This section summarizes the simulation domain for solvation number, $n$, temperature, $T$, and applied electric field magnitude, $E$, used to obtain dissociation statistics. The present study follows the same parameter space used in Smith et al. \cite{smith2026missing} for positive clusters and simulation region selection procedure. In contrast, the present work explicitly includes negative solvated ion clusters and therefore reports separate temperature ranges for [\ce{EMI-BF4}]$_n$\ce{EMI+} and [\ce{EMI-BF4}]$_n$\ce{BF4-} clusters.

Available measurements and plume-scale models indicate that ion clusters can depart the emission zone with internal energies as high as $3000$ K and cool rapidly toward the hundreds of Kelvin range within microseconds of flight \cite{prince2019solvated, miller2019characterization, miller2020measurement, nuwal2021multiscale, lyne2024inferring}. Therefore, prior MD campaigns \cite{prince2015molecular, schroeder2023inferring} span broad thermal windows to cover both near-emission and downstream conditions. Here we target the subset of that window where thermal dissociation becomes observable on nanosecond timescales across a large fraction of trajectories, while still overlapping the early flight temperatures inferred for electrospray beams \cite{miller2020measurement, nuwal2021multiscale, lyne2024inferring}. The local electric field near the Taylor cone apex is commonly estimated to be on the order of $10^9$~\unit{\volt/\m}, with fields decreasing by orders of magnitude through extraction optics and in the expanding plume \cite{chiu2007vacuum, coffman2019electrohydrodynamics, gallud2022emission, hampl2022comparison, magnani2023modelling, smith2024propagating, takagi2024simple}. To span both weak field and strong field limits, we apply uniform external fields from $10^6$ to $10^9$~\unit{\volt/\m}.

\begin{figure*}[htb]
    \centering
    \includegraphics[width=0.99\linewidth]{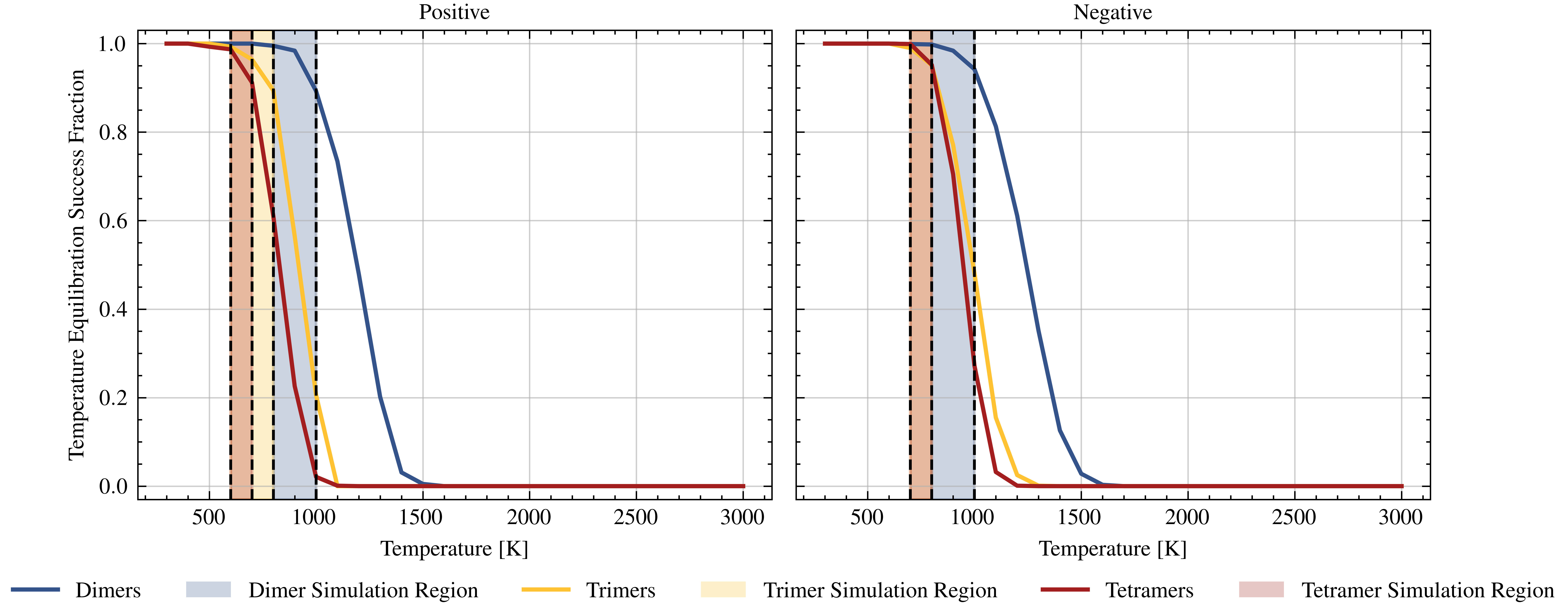}
    \caption{Left: Ion cluster temperature equilibration success fractions for \ce{[EMI-BF4]}\ce{EMI+} dimers, [\ce{EMI-BF4}]$_2$\ce{EMI+} trimers, and [\ce{EMI-BF4}]$_3$\ce{EMI+} tetramers at $E = 10^9$ \unit{\volt/\m}. Right: Ion cluster temperature equilibration success fractions for \ce{[EMI-BF4]}\ce{BF4-} dimers, [\ce{EMI-BF4}]$_2$\ce{BF4-} trimers, and [\ce{EMI-BF4}]$_3$\ce{BF4-} tetramers at $E = 10^9$ \unit{\volt/\m}.}
    \label{fig:temperature_equilibration_success}
\end{figure*}

In this study, cluster sizes are restricted to $n \le 3$. Beam diagnostics in the PIR consistently report that current is carried predominantly by bare ions and small solvated clusters, with larger solvation occurring as a low population tail \cite{lozano2005ionic, jia2022quantification, ulibarri2025direct}. The focus on $n = 1$ to $n = 3$ also aligns with the sensitivity of inferred mass flow to prompt dissociation of small clusters \cite{iribarne1976evaporation, loscertales1995experiments, hogan2010ion, smith2026missing}. Each \((n,T,E)\) node is sampled with \(N = 1000\) independent microcanonical trajectories. The upper limit of each temperature set is selected by requiring that at least $90\%$ of independently thermalized initial conditions remain intact at the end of the equilibration stage. Figure~\ref{fig:temperature_equilibration_success} reports the equilibration success fraction evaluated at $E = 10^9$~\unit{\volt/\m} as a conservative screen. The lower end of each temperature set is dictated by computational tractability such that at sufficiently low $T$, dissociation becomes too rare to capture with brute-force trajectories over practical wall-clock times. The resulting parameter space for positive and negative clusters are listed in Tables~\ref{tab:pos_parameter_space} and \ref{tab:neg_parameter_space}. 

\begin{table}[ht]
    \caption{\label{tab:pos_parameter_space}Parameter space explored in this study for positive \ce{EMI-BF4} clusters.
    Each \((n,T,E)\) node is sampled with \(N = 1000\) independent trajectories.}
    \begin{ruledtabular}
    \begin{tabular}{lcc}
    Solvation Number &
    Temperature [K] &
    Electric Field $^{\mathrm{a}}$ [\unit{\volt/\m}]  \\
    \hline \\ [-1.75ex]
    1 & 800, 900, 1000 & \(10^{6}\) – \(10^{9}\) \\
    2 & 600, 700, 800  & \(10^{6}\) – \(10^{9}\) \\
    3 & 600, 650, 700  & \(10^{6}\) – \(10^{9}\) \\
    \end{tabular}
    \end{ruledtabular}
    \begin{flushleft}
    \footnotesize
    \(^{\mathrm{a}}\) Nine discrete values:
    \(10^{6}, 5\times10^{6}, 10^{7}, 5\times10^{7}, 10^{8}, 2.5\times10^{8}, 5\times10^{8}, 7.5\times10^{8}, 10^{9} \) \unit{\volt/\m} .
    \end{flushleft}
\end{table}

\begin{table}[ht]
    \caption{\label{tab:neg_parameter_space}Parameter space explored in this study for negative \ce{EMI-BF4} clusters.
    Each \((n,T,E)\) node is sampled with \(N = 1000\) independent trajectories.}
    \begin{ruledtabular}
    \begin{tabular}{lcc}
    Solvation Number &
    Temperature [K] &
    Electric Field $^{\mathrm{a}}$ [\unit{\volt/\m}]  \\
    \hline \\ [-1.75ex]
    1 & 800, 900, 1000 & \(10^{6}\) – \(10^{9}\) \\
    2 & 700, 750, 800  & \(10^{6}\) – \(10^{9}\) \\
    3 & 700, 750, 800  & \(10^{6}\) – \(10^{9}\) \\
    \end{tabular}
    \end{ruledtabular}
    \begin{flushleft}
    \footnotesize
    \(^{\mathrm{a}}\) Nine discrete values:
    \(10^{6}, 5\times10^{6}, 10^{7}, 5\times10^{7}, 10^{8}, 2.5\times10^{8}, 5\times10^{8}, 7.5\times10^{8}, 10^{9} \) 
    \end{flushleft}
\end{table}

\subsubsection{Simulation Parameterization and Production Protocol}

Microcanonical dissociation trajectories were performed in the Large-scale Atomic Molecular Massively Parallel Simulator (LAMMPS) \cite{thompson2022lammps} using its parallel short-range force evaluation and time integration framework \cite{plimpton1995fast}. The present study leverages the same fixed-charge CHARMM General Force Field (CGenFF) \cite{mackerell1998all, vanommeslaeghe2010charmm} all-atom force field used in Smith et al. \cite{smith2026missing} for both positive and negative polarities. Bonded terms include harmonic bond stretches, angle bends, and improper torsions, while dihedral torsions follow the standard CHARMM cosine series. Non-bonded interactions are given by a 12-6 Lennard-Jones (LJ) potential and a Coulomb term between atom-centered partial charges. The resulting energy is given by

\begin{equation}
    \begin{aligned}
    U_{\mathrm{tot}} =
    & \sum_{\text{bonds}} k_r\,(r-r_0)^2
    +\sum_{\text{angles}} k_\theta\,(\theta-\theta_0)^2 \\[2pt]
    \quad
    +&\sum_{\text{dihedrals}} k_\phi\,[1+\cos(n\phi-\lambda)] \\[4pt]
    \quad+&\sum_{\text{impropers}} k_\psi\,(\psi-\psi_0)^2 \\[4pt]
    +&\quad\sum_{i<j}\Bigl[
    4\varepsilon_{ij}\!\left(\tfrac{\sigma_{ij}}{r_{ij}}\right)^{\!12}
    -4\varepsilon_{ij}\!\left(\tfrac{\sigma_{ij}}{r_{ij}}\right)^{\!6}
    +\tfrac{q_i q_j}{4\pi\varepsilon_0\,r_{ij}}
    \Bigr].
    \end{aligned}
    \label{equation:total_energy}
\end{equation}

LJ parameters for unlike atom pairs use Lorentz-Berthelot mixing \cite{lorentz1881ueber, berthelot1898melange} and standard CHARMM 1-4 scaling is applied \cite{mackerell1998all, vanommeslaeghe2010charmm}. LJ interactions are smoothly switched from 10~{\AA} to 12~{\AA}, and Coulomb interactions are truncated at 1~\unit{\micro\meter}. For each \(n,T,E\) node in Tables~\ref{tab:pos_parameter_space} and \ref{tab:neg_parameter_space}, $1000$ independent microcanonical, non-periodic shrink-wrapped trajectories in both polarities, following prior electrospray MD fragmentation workflows \cite{prince2019solvated, nuwal2021multiscale, schroeder2023inferring} and the detailed implementation described in Smith et al. \cite{smith2026missing}. Each trajectory is initialized from an energy-minimized relaxed cluster, assigned Maxwell-Boltzmann velocities, equilibrated to the target temperature with a Nos\'e-Hoover thermostat and weak Langevin damping to suppress residual rigid-body rotation. Production dissociation trajectories are then propagated microcanonically with a timestep of $0.5$~\unit{\femto\second} under a uniform static electric field is applied along the $\hat{x}$ direction. Dissociation is detected from the first loss of connectivity of a distance graph constructed from instantaenous atomic coordinates using a conservation bond cutoff of $d_b = 5$~{\AA}. The corresponding index defines $k_{\mathrm{frag}}$ and the lifetime $\tau = \left(k_{\mathrm{frag}}-k_0\right)\Delta t$, after which trajectories are continued until the fragment centers of mass separate beyond $d_f = 40$~{\AA} to confirm irreversible dissociation \cite{coles2013investigating} 

\begin{figure*}[ht!]
    \centering
    \includegraphics[width=0.85\linewidth]{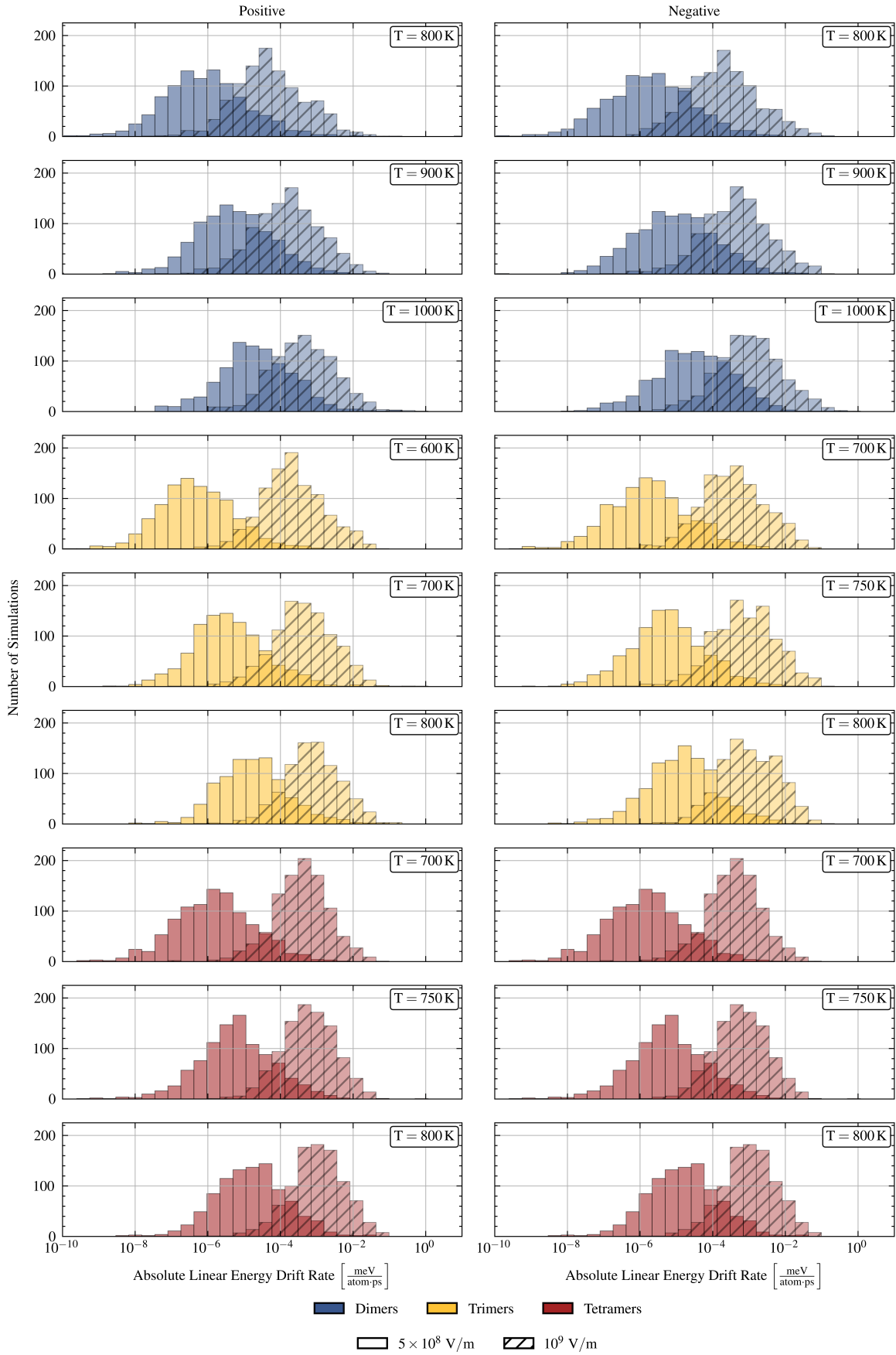}
    \caption{Left: Ensemble distributions of microcanonical absolute linear energy drift rates for positive [\ce{EMI-BF4}]\ce{EMI+} dimers, [\ce{EMI-BF4}]$_2$\ce{EMI+} trimers, and [\ce{EMI-BF4}]$_3$\ce{EMI+} tetramers at $E = 5 \times 10^8$ and $E = 10^9$ \unit{V}/\unit{m}. Right: Ensemble distributions of microcanonical absolute linear energy drift rates for negative [\ce{EMI-BF4}]\ce{BF4-} dimers, [\ce{EMI-BF4}]$_2$\ce{BF4-} trimers, and [\ce{EMI-BF4}]$_3$\ce{BF4-} tetramers at $E = 5 \times 10^8$ and $E = 10^9$ \unit{V}/\unit{m}.}
    \label{fig:energy_drift}
\end{figure*}

Energy conservation during the microcanonical production runs is verified by fitting a line to the total energy time series for each trajectory and reporting the absolute drift rate per atom. Figures~\ref{fig:energy_drift} depict ensemble-scale drift rate distributions for all clusters studied in this work, indicating stable integration over the dissociation timescales considered due to the strong concentration at very small values. 

\subsubsection{Dissociation Lifetime Parametrization Procedure} \label{sec:lifetime_model}

The resultant dissociation lifetimes span multiple orders of magnitude across the electrospray thruster operating field range, motivating the need for a compact model that augments a thermal dissociation free-energy with a field-assisted barrier reduction. Assuming first-order dissociation, we interpret the ensemble mean lifetime using a transition theory state expression in which the applied electric field lowers the effective activation free energy \cite{prince2017combined, miller2020measurement, schroeder2023inferring}. To accommodate both weak and strong field limits encountered in ionic liquid electrosprays, the barrier-lowering work $G(E)$ is represented as a smooth interpolation between the image potential model (IPM) commonly used in ion evaporation and field-assisted emission applications \cite{iribarne1976evaporation, loscertales1995experiments,coffman2019electrohydrodynamics} and the dipole potential model (DPM) that captures finite charge redistribution along the dissociation coordinate as described in Smith et al. \cite{smith2026missing}:

\begin{subequations}
    \begin{align}
    G_{\mathrm{IPM}}(E)
      &= \sqrt{\frac{q^3 E}{4\pi\epsilon_0}},
      \label{equation:ipm}\\
    G_{\mathrm{DPM}}(E)
      &= qEx + \frac{q^2}{4\pi\epsilon_0}
         \left(\frac{1}{x}-\frac{1}{x+d}\right),
      \label{equation:dpm}\\
    \alpha(E)
      &= \frac{1}{1+(E/E_c)^2},
      \label{equation:mixing}\\
    G(E)
      &= [1-\alpha(E)]G_{\mathrm{IPM}}(E) \notag\\
      &\quad +\alpha(E)G_{\mathrm{DPM}}(E)
              -G_{\mathrm{DPM}}(0).
      \label{equation:g_model}
    \end{align}
\end{subequations}

where $q =|Z|e$ denotes the cluster charge magnitude and $E$ is the applied field magnitude. The parameter $x$ is the characteristic separation along the dissociation coordinate, $d$ is the effective dipole length which is fixed to $1$~\unit{\nm} \cite{schroeder2021numerical}, and $E_c$ controls the crossover between the two limiting forms. The resultant lifetime model is described in Equation~\ref{eq:lifetime}:

\begin{equation}
    \begin{aligned}
        \ln{\tau\left(E,T\right)} = \frac{\Delta G_n^0 - G(E)}{k_B T} + \ln{\left(\frac{h}{k_B T}\right)}
    \end{aligned}
    \label{eq:lifetime}
\end{equation}

with $\Delta G_n^0$ representing the the zero-field dissociation for solvation number $n$. For each $n$ and polarity combination, teh parameters, $\Delta G_n^0$, $x$, and $E_c$ are obtained by nonlinear regression of the simulated $\ln{\tau\left(E,T\right)}$ MD dataset over the sampled parameter grid. Equation~\ref{eq:lifetime} reduces to the standard thermal Arrhenius dependence in the $E\!\to\!0$ limit, while at the high field limit the barrier-lowering term dominants and reproduces monotonic reduction of lifetime with increasing $E$.

\section{Results}
\subsection{Electric Field and Temperature Dependent Dissociation Lifetimes} \label{sec:md_results}

\begin{figure*}[ht!]
    \centering
    \includegraphics[width=0.90\linewidth]{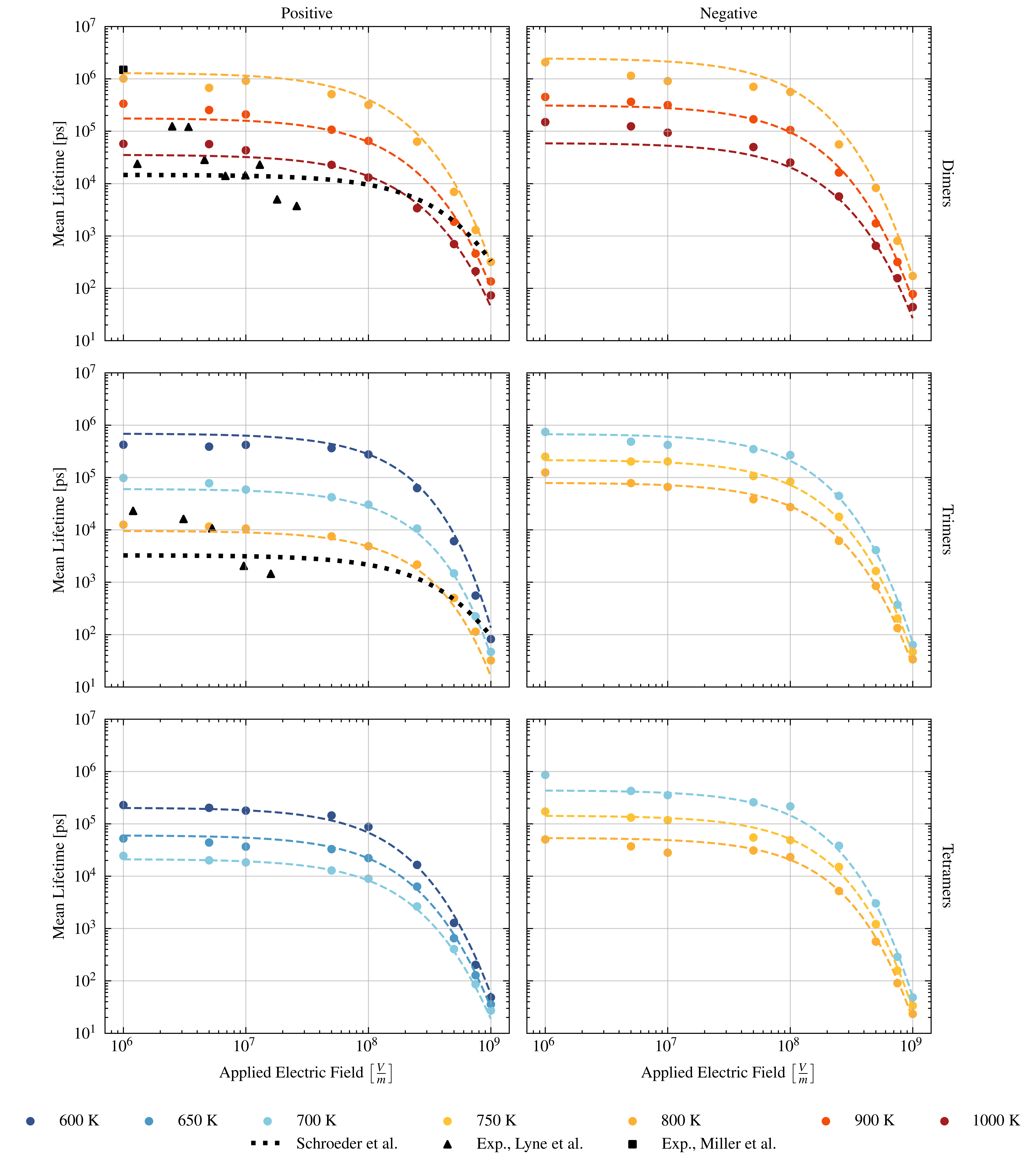}
    \caption{Left: Ion cluster mean lifetimes for [\ce{EMI-BF4}]\ce{EMI+} dimers, [\ce{EMI-BF4}]$_2$\ce{EMI+} trimers, and [\ce{EMI-BF4}]$_3$\ce{EMI+} tetramers. Right: Ion cluster mean lifetimes for [\ce{EMI-BF4}]\ce{BF4-} dimers, [\ce{EMI-BF4}]$_2$\ce{BF4-} trimers, and [\ce{EMI-BF4}]$_3$\ce{BF4-}.}
    \label{fig:dissociation_lifetimes}
\end{figure*}

Ion cluster mean lifetimes for positive and negative \ce{EMI-BF4} dimers, trimers, and tetramers are shown in Figure~\ref{fig:dissociation_lifetimes}. Field-dependent exponential fits are superimposed as visual guides with fitted parameters for positive \ce{EMI-BF4} clusters reported in Table~\ref{tab:fitted_params} and for negative \ce{EMI-BF4} clusters reported in Table~\ref{tab:negative_fitted_params}.

\begin{table}[ht]
    \caption{\label{tab:fitted_params}Solvation number dependent fits for positive \ce{EMI-BF4} clusters collected in this study.}
    \begin{ruledtabular}
    \begin{tabular}{lccc}
    Solvation Number &
    $\Delta G_n^{\ddagger}$ [\unit{\electronvolt}] &
    $E_c$ [\unit{\volt/\m}] &
    $x$  [\unit{\angstrom}] \\ 
    \hline \\ [-1.75ex]
    1 & $1.164$ & $8.979 \times 10^{8}$ & $8.811$ \\ 
    2 & $0.826$ & $6.198 \times 10^{9}$ & $4.751$ \\
    3 & $0.762$ & $2.086 \times 10^{9}$ & $5.946$ \\
    \end{tabular}
    \end{ruledtabular}
    \begin{flushleft}
    \end{flushleft}
\end{table}

\begin{table}[ht]
    \caption{\label{tab:negative_fitted_params}Solvation number dependent fits for negative \ce{EMI-BF4} clusters collected in this study.}
    \begin{ruledtabular}
    \begin{tabular}{lccc}
    Solvation Number &
    $\Delta G_n^{\ddagger}$ [\unit{\electronvolt}] &
    $E_c$ [\unit{\volt/\m}] &
    $x$  [\unit{\angstrom}] \\ 
    \hline \\ [-1.75ex]
    1 & $1.209$ & $7.278 \times 10^{8}$ & $10.033$ \\ 
    2 & $0.972$ & $1.551 \times 10^{9}$ & $7.449$ \\
    3 & $0.945$ & $2.055 \times 10^{9}$ & $6.877$ \\
    \end{tabular}
    \end{ruledtabular}
    \begin{flushleft}
    \end{flushleft}
\end{table}

\subsubsection{Polarity Dependent Dimer Dissociation Lifetimes}

The mean lifetime for positive \ce{EMI-BF4} dimers remains field-independent up to $E \approx 10^7$ \unit{\volt/\m}, corresponding to plateaus of $1.3 \times 10^6$, $1.8 \times 10^5$, and $3.6 \times 10^4$ \unit{\ps} at $800$, $900$, and $1000$ K. Negative \ce{EMI-BF4} dimers exhibit the same temperature-controlled behavior over this range with slightly longer plateau lifetimes of $2.4 \times 10^6$, $3.0 \times 10^5$, and $6.0 \times 10^4$ \unit{ps} at $800$, $900$, and $1000$ K. This behavior mirrors previous MD studies in which the Coloumbic work, $qE\Delta r$, is smaller relative to the internal thermal energy, $k_B T$, of the ion cluster and therefore unable to appreciably lower the activation barrier \cite{schroeder2021numerical, schroeder2023inferring}. At $10^8$ \unit{\volt/\m} a clear inflection point appears such that mean lifetimes begin to fall by roughly one order of magnitude for every order of magnitude increase in field strength, signaling a field-controlled regime analogous to that reported by Schroeder et al \cite{schroeder2023inferring} for imidazolium clusters. At the highest field investigated, $10^9$ \unit{\volt/\m}, positive dimer clusters survive only $323$, $110$, and $46$ \unit{\ps} for $800$, $900$, and $1000$ K, while negative dimers survive $172$, $77$, and $44$ \unit{ps}, demonstrating that the effect of the applied electric field dominates over thermal activation in the strong field regime \cite{kapranov2013stochastic, su2023fragmentation}. The modestly increased stability of negative dimers in the low field regime is consistent with field-free RPA measurements that report microsecond timescale lifetimes in both polarities with a slightly longer mean lifetime for negative \ce{EMI-BF4} dimers \cite{miller2020measurement}.

\subsubsection{Polarity Dependent Trimer Dissociation Lifetimes}

The mean lifetimes of positive and negative \ce{EMI-BF4} trimers depict the same qualitative sequence in which the the lifetimes remain field-independent up to $E \approx 2.5 \times 10^7$ \unit{\volt/\m}. This field value corresponds to plateaus of $6.8 \times 10^5$, $6.0 \times 10^4$, and $9.4 \times 10^3$ \unit{\ps} at $600$, $700$, and $800$ K for positive trimers and $7 \times 10^5$, $2 \times 10^5$, and $8 \times 10^4$ \unit{\ps} at $700$, $750$, and $800$ K. These plateaus indicated that negative trimers possess enhanced stability relative to positive trimers. For any fixed applied electric field and temperature, the trimer lifetime is $0.5$ to $1.5$ orders of magnitude lower than the dimer lifetime, which is consistent with the size‑dependent destabilization observed in other ionic liquid clusters \cite{viitanen2008experimental, hsu2024electric}. Once the applied electric field is larger than $10^8$ \unit{\volt/\m}, the positive trimer lifetime collapses to the dimer envelope of $0.277$, $0.031$, and $0.005$ \unit{\us} for $600$, $700$, and $800$ K and $0.269$, $0.083$, and $0.027$ for $700$, $750$, and $800$ K, indicating that the dissociation rate is dictated exclusively by the electrostatic work exerted during charge separation.

\subsubsection{Polarity Dependent Tetramer Dissociation Lifetimes}

The mean lifetimes of positive and negative \ce{EMI-BF4} tetramers display the same correlation observed for dimers and trimers, consisting of a low-field, temperature-controlled regime followed by a rapid transition to field-assisted dissociation at high applied fields. In the temperature-controlled regime, positive [\ce{EMI-BF4}]$_3$\ce{EMI+} tetramers remain approximately field-independent up to $E \approx 2.5 \times 10^7$~\unit{\volt/\m}, corresponding to plateau lifetimes of $2.3 \times 10^5$, $5.2 \times 10^4$, and $2.4 \times 10^4$~\unit{\ps} at $600$, $650$, and $700$ K. Negative [\ce{EMI-BF4}]$_3$\ce{BF4-} tetramers exhibit the same temperature-controlled behavior over this regime, but with systematically longer lifetimes. The corresponding plateaus are $8.5 \times 10^5$, $1.7 \times 10^5$, and $5.0 \times 10^4$~\unit{\ps} at $700$, $750$, and $800$ K. The increased low field lifetimes of negative tetramers indicate greater stability relative to positive tetramers, which is consistent with the polarity-dependent stability trends observed for \ce{EMI-BF4} trimers. At any fixed applied electric field and temperature, tetramers are the least stable of the clusters studied in this work with mean lifetimes reduced relative to trimers by $0.3$ to $0.7$ orders of magnitude in the temperature-controlled regime. 

\subsection{Dissociation Pathway Mechanisms}

\subsubsection{Polarity Dependent Trimer Dissociation Branching Mechanisms}

\begin{figure*}[ht!]
    \centering
    \includegraphics[width=0.90\linewidth]{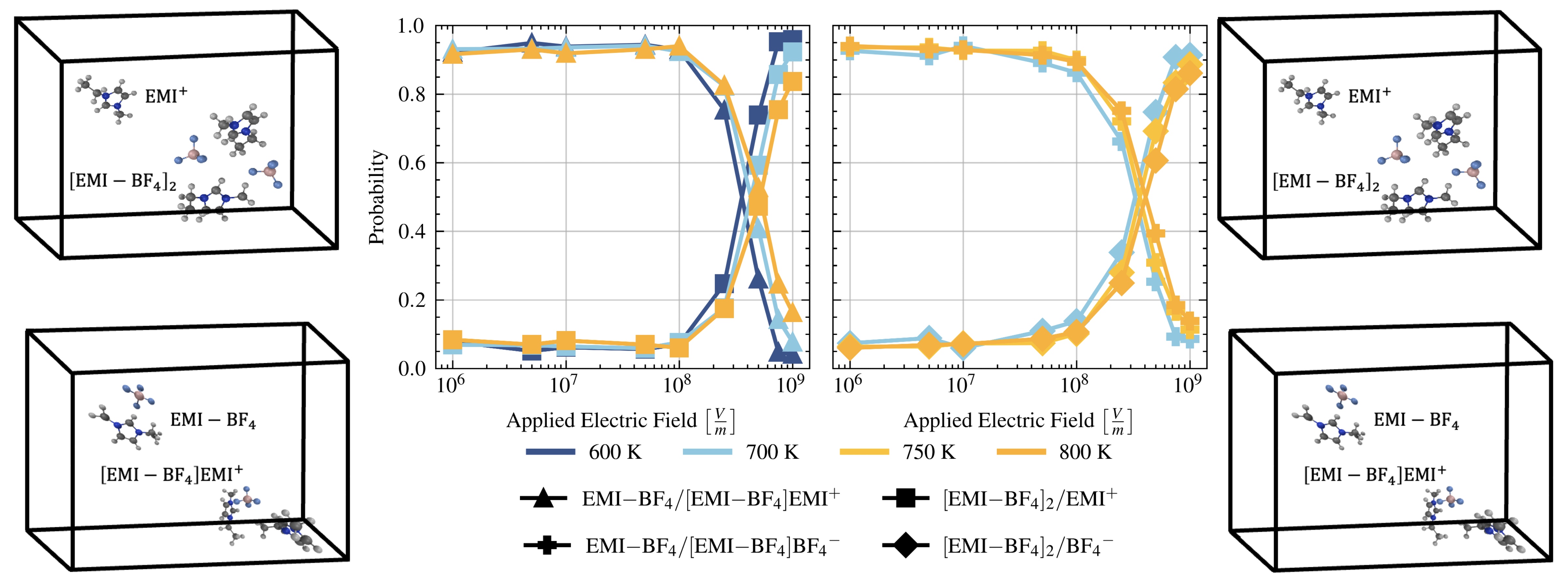}
    \caption{Left: Ion cluster fragmentation pathways for [\ce{EMI-BF4}]$_2$\ce{EMI+} trimers at $600$, $700$, and $800$ K. Right: Ion cluster fragmentation pathways for [\ce{EMI-BF4}]$_2$\ce{BF4-} trimers at $700$, $750$, and $800$ K. Increasing the applied electric field shifts the dominant trimer dissociation pathway from single neutral ion-pair evaporation to charged core ion ejection in both polarities.}
    \label{fig:trimer_pathways}
\end{figure*}

Figure~\ref{fig:trimer_pathways} tracks the fragmentation pathways of positive and negative \ce{EMI-BF4} trimers across the simulated temperature range. For both polarities, the trimer dissociation is well-described by competition between two primary outcomes. The first pathway is represented by evaporation of a single neutral ion pair, \ce{EMI-BF4}, which reduces the solvation number by one and produces a charged dimer. The second pathway is described by field-driven ejection of teh charged core ion, producing an isolated monomer and leaving behind a double neutral cluster, [\ce{EMI-BF4}]$_2$. In positive polarity, Pathway~I corresponds to \ce{EMI-BF4} + [\ce{EMI-BF4}]\ce{EMI+}, whereas Pathway~II corresponds to \ce{EMI+} + [\ce{EMI-BF4}]$_2$. In negative polarity, the analogous channels are \ce{EMI-BF4} + [\ce{EMI-BF4}]\ce{BF4-} and \ce{BF4-} + [\ce{EMI-BF4}]$_2$. At low applied electric fields, the neutral pair evaporation channel in both polarities with probabilities exceeding $90\%$ over the sampled temperatures and fields below $10^8$~\unit{\volt/\m}. Over the same low-field range, charged ion core ejection remains a minor channel, typically account for only a few percent of events. Above a temperature-independent critical applied electric field greater than $10^8$ \unit{\volt/\m}, the charged core ejection pathways rises sharply and becomes dominant in the strong field regime. At applied electric fields greater than $10^8 {V}/{m}$, this pathway transition occurs on the same order as surface fields associated with ion evaporation and onset of monomer emission from ILs in charged droplets, larger clusters, and stressed menisci \cite{iribarne1976evaporation, coffman2019electrohydrodynamics, bhakyapaibul2025electric}. This observed mechanism is consistent with field-induced alignment of the cluster dipole, which preferentially stretches the cation-anion contact along the field direction that ultimately breaks \cite{kapranov2013stochastic, su2023fragmentation}. Notably, the critical field required to trigger this shift in the dominant dissociation channel is comparable to the characteristic field scale for onset of ion extract at the emitter apex, suggesting that strong field environments in the near-source acceleration region can fundamentally alter not only the dissociation rate but also the dissociation pathway for solvated clusters.



\subsubsection{Polarity Dependent Tetramer Dissociation Branching Mechanisms}

\begin{figure*}[ht!]
    \centering
    \includegraphics[width=0.90\linewidth]{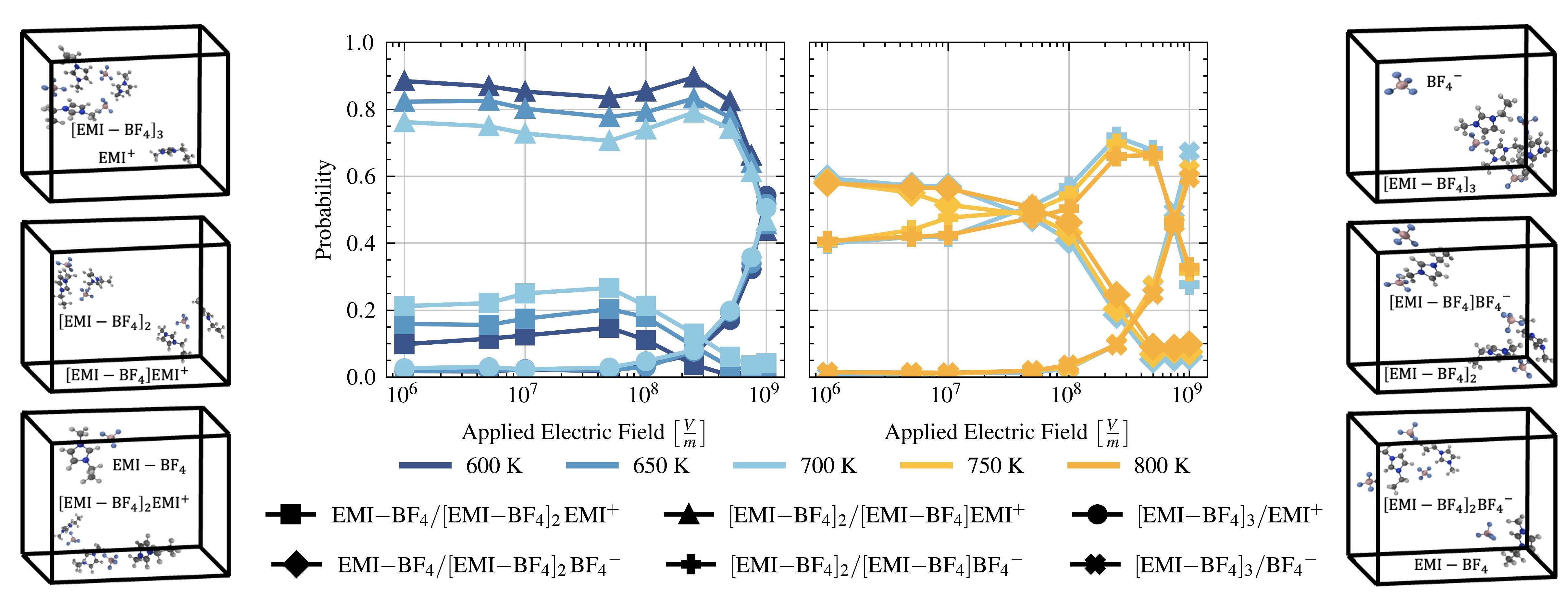}
    \caption{Left: Ion cluster fragmentation pathways for [\ce{EMI-BF4}]$_3$\ce{EMI+} tetramers at $600$, $650$, and $700$ K. Right: Ion cluster fragmentation pathways for [\ce{EMI-BF4}]$_3$\ce{BF4-} tetramers at $700$, $750$, and $800$ K. Increasing the applied electric field promotes charged core ion ejection in both polarities, while tetramer branching between single neutral ion-pair evaporation and double neutral cluster emission at lower fields depends on polarity.}
    \label{fig:tetramer_pathways}
\end{figure*}

Figure~\ref{fig:tetramer_pathways} tracks the dissociation pathway branching for positive and negative \ce{EMI-BF4} tetramers across the simulated temperature range. In contrast to the dissociation pathways mapped for \ce{EMI-BF4} trimers, tetramer breakup consistently exhibits three competing outcomes that differ by the number of neutral pair units removed from the parent cluster. Pathway~I corresponds to evaporation of a single neutral ion pair, \ce{EMI-BF4}, producing a charged trimer and reducing the solvation number by one. In positive polarity, this channel is \ce{EMI-BF4} + [\ce{EMI-BF4}]$_2$\ce{EMI+}, and, in negative polarity, it is \ce{EMI-BF4} + [\ce{EMI-BF4}]$_2$\ce{BF4-}. Pathway~II corresponds to emission of a double neutral cluster, [\ce{EMI-BF4}]$_2$, producing a charged dimer and reducing the solvation number by two. In positive polarity, this channel is [\ce{EMI-BF4}]$_2$ + [\ce{EMI-BF4}]\ce{EMI+}, and, in negative polarity, it is [\ce{EMI-BF4}]$_2$ + [\ce{EMI-BF4}]\ce{BF4-}. Pathway~III corresponds to field-driven ejection of the charged core ion, producing an isolated monomer and leaving a triple neutral cluster, [\ce{EMI-BF4}]$_3$, and reducing the solvation number by three. In positive polarity, this channel is \ce{EMI+} + [\ce{EMI-BF4}]$_3$, and, in negative polarity, it is \ce{BF4-} + [\ce{EMI-BF4}]$_3$.

At low applied electric fields less than $10^8$~\unit{\volt/\m}, tetramer dissociation is dominated by pathways that retain a solvated charged product, Pathways~I and II, while direct core ion ejection remains negligible in both polarities. However, the low field branching differs markedly between positive and negative tetramers. Positive tetramers preferentially dissociate by emitting a double neutral cluster with probabilities in the range of $75\%$  to $90\%$ over the simulated temperatures, whereas single neutral pair evaporation remains a secondary channel at the level of $10\%$ to $30\%$. Notably, positive tetramers show a clear temperature dependence in this low field regime such that as temperature increases, the probability of Pathway~I increases and the probability of Pathway~II decreases. This correlation indicates that thermal activation promotes stepwise neutral pair evaporation relative to the emission of a bound neutral dimer. In contrast, negative tetramers denote little to no temperature dependence in pathway probability across the full temperature range. In the low field regime, negative tetramers show a comparatively balanced partitioning between Pathways ~I and II with single neutral pair evaporation accounting for $55\%$ to $60\%$ of events and double neutral emission accounting for $40\%$ to $45\%$. 

As the applied field increases into the intermediate range, between $10^8$ to $5 \times 10^8$~\unit{\volt/\m}, field polarization begins to reshape the preferred cleavage topology for tetramers, particularly in negative polarity. For negative tetramers, a pronounced crossover occurs near $10^8$~\unit{\volt/\m}, where Pathway~II becomes dominant and reaches peak probabilities fo $65\%$ to $75\%$ across the temperature range, while Pathway~I is suppressed to approximately $20\%$. Over the same applied field range, positive tetramers remain dominated by Pathway~II, but the temperature dependent persists such that higher temperatures maintain a larger contribution of single neutral evaporation and delay the extent to which the field concentrates dissociation into higher count neutral clusters. Across both polarities, these trends are consistent with the applied field preferentially stabilizing dissociation topologies that preserve the most strongly bound solvation environment around the charged product while weakening the outermost ion-pair contacts along the field direction \cite{schroeder2023inferring, su2023fragmentation}.

In the strong field regime, direct ejection of the charged core ion rises sharply in both polarities becoming the major dissociation channel. For positive tetramers, Pathway~III increases rapidly above a few $10^8$~\unit{\volt/\m} and approaches probabilities on the order of $45\%$ to $55\%$ by $10^9$~\unit{\volt/\m}, becoming comparable to Pathway~II while Pathway~I becomes negligible. For negative tetramers, the rise of Pathway~III is even more pronounced such that this pathway reaches probabilities of $60\%$ to $70\%$ while Pathway~II correspondingly decreases to $30\%$ and Pathway~I remains negligible. The onset and rapid growth of the core ion ejection pathway occurs at electric field magnitudes comparable to those associated with ion evaporation and monomer emission from highly stressed IL surfaces and charged droplets as observed for \ce{EMI-BF4} trimers \cite{iribarne1976evaporation, coffman2019electrohydrodynamics, bhakyapaibul2025electric}. The polarity dependence indicates that while strong fields universally promote core ion ejection, the competition between single and double neutral pathways at lower fields remains sensitive to the detailed charge distribution and bond topology of the parent tetramer.



\section{Discussion}

The MD study presented in Section~\ref{sec:md_results} provides a pathway-resolved kinetic description of \ce{EMI-BF4} solvated ion clusters under conditions representative of the electrospray acceleration environment. Across both polarities and solvation numbers $n = 1$ to $n = 3$, the principal outcome is a consistent transition from a low field, temperature-controlled dissociation regime to a strong field regime in which electrostatic work dominates the barrier to fragmentation. This structure provides a physically interpretable framework for connecting atomistic fragmentation dynamics to the near-field electrospray environment \cite{nuwal2021multiscale, petro_2022, schroeder2023inferring, smith2024propagating}.

The primary question addressed in this work is whether cluster dissociation during acceleration can be treated as purely thermally activated, or whether field-assisted barrier lowering must be included explicitly. Prior experimental and computational efforts strongly support first-order survival statistics in weak field regions downstream of the extractor with microsecond timescale lifetimes that vary rapidly with internal temperature \cite{miller2020measurement}. By contrast, the near field acceleration region is characterized by rapidly varying fields and short residence times, which motives the used of MD to inform field-assisted rate models in coupled plume simulations \cite{nuwal2021multiscale, petro_2022, schroeder2023inferring, smith2024propagating}. The present MD results reinforce this picture by resolving a regime boundary such that for all solvation numbers considered, mean lifetimes remain field-independent at low fields and then collapse rapidly once the applied field increases to $10^8$~\unit{\volt/\m}. This qualitative correlation agrees with prior MD approaches that identify a crossover from temperature-controlled dissociation to dissociation dominated by applied fields for \ce{EMI-BF4} clusters \cite{schroeder2023inferring}. The implication of the present results for multiscale modeling is that treating dissociation with a field-free rate constant in the acceleration region will generally underpredict prompt fragmentation and misrepresent the timing and location of dissociation events, particularly for higher solvation number clusters that exhibit shorter lifetimes at the same temperature and applied field. 

The polarity dependence observed in the presented lifetime and pathway branching results further indicate that a single, polarity independent dissociation model is generally insufficient for quantitative plume interpretation. In the low field regime, negative clusters are systematically more stable than their positive counterparts. This trend is consistent with field-free measurements that report similar timescale lifetimes in both polarities with modestly increased stability for \ce{EMI-BF4} dimers \cite{miller2020measurement}. However, in the strong field limit, lifetimes in both polarities converge toward nanosecond timescale lifetimes, indicating that once $qE\Delta r$ becomes large relative to $k_BT$, barrier lowering dominates and polarity-dependent energetic differences become secondary. Practically, this suggests that polarity dependence is most consequential in the temperature-controlled regime, whereas strong field dissociation may admit a more universal description such that the applied field contribution controls the kinetics.

The solvation number dependence extracted here is likewise aligned with established observations that cluster stability decreases with increasing cluster size for ILs in the gas phase. Sequential neutral pair evaporation has been identified in numerous studies \cite{hogan2010ion, prince2015molecular} as the dominant field-free dissociation mechanism in isolated IL clusters. In that context, larger parent clusters exhibit lower effective stability because outer solvation units are more weakly bounded, meaning additional fragmentation topologies become accessible at the number of ion pairs increases. The MD lifetimes presented in this work preserve this ordering in the temperature-controlled regime, where tetramers are least stable, followed by trimers, and then dimers. Importantly, the present results show that this ordering is not merely an intrinsic thermal effect but persists into the intermediate field regime that is most relevant for electrospray acceleration. As the applied field strength increases, all clusters eventually transition into a regime where dissociation rates are dictated primarily by electrostatic work, and the solvation number dependence becomes compressed. From the standpoint of plume transport, this compression implies that, in sufficiently strong field, differences in initial solvation number may exert a smaller influence on where dissociation occurs than would be inferred from field-free rates alone. 

The presented dissociation branching results underscores that applied fields alter not only how fast clusters dissociation, but also how they dissociate, with direct consequences for the downstream charge and neutral densities. For trimers, dissociation across both polarities can be reduced to competition between neutral pair evaporation and charged core ion ejection with a sharp field-driven transition near $10^8$~\unit{\volt/\m}. The emergence of core ion ejection at high fields is qualitatively consistent with classical ion evaporation concepts, where strong fields lower the effective free-energy barrier and promote charge separation pathway that are otherwise suppressed \cite{iribarne1976evaporation}. For tetramers, three pathways are consistently observed: single neutral evaporation, double neutral emission, and core ion ejection leaving a triple neutral cluster. The additional channel observed at $n = 3$ emphasizes that higher solvation number clusters cannot be represented accurately by a single sequential evaporation mechanism in the acceleration. Instead, multiple neutral loss channels compete and the dominant pathway can change with field magnitude. The polarity dependence tetramer branching is also notable such that negative tetramers show weak temperature dependence in pathway probabilities comparatively to positive clusters. One plausible interpretation is that the internal structural anisotropy and charge distribution of positive clusters enables thermally mediated access to competing cleavage topologies, while negative clusters exhibit a more field-driven partitioning. These results motivate that pathway-resolved modeling for $n \ge 3$ because temperature-averaged branching fractions would mask physically meaningful shifts in the charged product distribution.

The mechanistic findings presented in this work have direct implications for the interpretations of downstream diagnostic and for the fidelity of multiscale electrospray plume simulations. RPA and TOF diagnostics primarily sample the charged population, yet the dissociation pathways quantified here produce substantial neutral mass that is invisible to the detectors used in these diagnostics \cite{schroeder2023inferring, lyne2024inferring}. In particular, triple neutral branching for tetramers and double neutral emission for both trimers and tetramers, redistribute mass into neutral clusters that may carry a significant fraction of the total momentum and energy but produce no direct current signal. Consequently, diagnostic inversion methods that assume a limited set of fragmentation channels can systematically misattribute charge fractions and infer biased propellant utilization when fragmentation occurs within the measurement region or upstream of it \cite{nuwal2021multiscale, petro_2022}. The present results enables a more complete forward model of charged and neutral production during acceleration and can therefore reduce model form uncertainty in diagnostic interpretation. More broadly, the combination of field-dependent lifetimes and pathway branching furnishes the minimum set of atomistic inputs needed to propagate metastable clusters in $N$-body acceleration solvers while preserving consistent fragmentation products and kinetics during the transient, strong field portion of flight.

Finally, the limitations of the present approach point to clear directions for future work. First, the MD framework employed here relies on a non-polarizable, fixed-charge force field. It is well known that fixed-charged IL models can misestimate cohesive energetics and ion-pair interaction strengths because they neglect electronic polarization and charge redistribution \cite{bedrov2019molecular}. While the present work partially mitigates this limitation by focusing on relative trends across temperature, applied field, and solvation number and by fitting reduced-order lifetime models to ensemble statistics, improved absolute accuracy may require polarizable force fields or ab initio methods \cite{laws2025ab}. Second, the applied electric fields in MD are spatially uniform, whereas realistic electrospray acceleration fields vary rapidly over \unit{\nm} to \unit{\micro\metre} length scales and are coupled to evolving cluster orientation and translation. Coupling the presented parameterizations for lifetime to self-consistent electrospray field solutions in plume simulations remains essential for predicting where in the acceleration region fragmentation is most probable \cite{nuwal2021multiscale, petro_2022, smith2024propagating}. Third, the accessible temperature range is bounded by equilibration success for larger clusters and by computational feasibility for long-lived, low temperature conditions. Extending the ensemble methodology to colder internal energies and to larger solvation numbers, while retaining statistically converged lifetimes and branching fractions will be important for modeling studies where clusters cool rapidly downstream or where higher $n$ clusters contribute non-negligibly to mass flow. Addressing these challenges will further strengthen the transferability of MD fragmentation kinetics across operation conditions and geometries. 

\section{Conclusions}

This work presents a pathway-resolved MD investigation of field-assisted thermal dissociation thermal dissociation of solvated \ce{EMI-BF4} ion clusters under conditions relevant to PIR operation. For dimers, trimers, and tetramers in both polarities, large trajectory ensembles were generated across temperatures and uniform electric fields representative of the electrospray thruster acceleration region. Fragmentation lifetimes and product identities were extracted using a connectivity-based fragmentation criterion that enables converged statistics across broad conditions. The resulting dataset addresses a persistent bottleneck in multiscale electrospray modeling, where plume transport solvers require transferable kinetic inputs that cannot be isolated cleanly from measurements along \cite{lozano2005ionic, zorzos2008use, nuwal2021multiscale, petro_2022, smith2024propagating}. These kinetics are uniquely positioned to supply required inputs for plume transport and diagnostic models. 

Across solvation numbers $n = 1$ to $n = 3$, the inferred mean lifetimes span multiple orders of magnitude over the sampled temperatures and fields. In the weak field limit, the ensemble recovers thermally activated dissociation that is compatible with first-order survival behavior inferred from field-free measurements downstream of the extractor \cite{miller2020measurement}. As fields increase into the near-source regime, lifetimes decrease steeply and reach sub-nanosecond values in the strongest fields sampled, bridging microsecond-scale downstream decay and nanosecond-scale stability implied by acceleration region measurement \cite{lyne2024inferring}. The polarity dependence is most pronounced where thermal activation dominates, whereas in the strongest fields the lifetimes of positive and negative clusters approach similar magnitudes as electrostatic work becomes the primary contribution.

The second outcome of the present work is a reduced-order parameterization of $\tau_n\left(E,T\right)$ grounded in transition state theory and field-assisted barrier lowering. By representing the electrostatic work along the separation coordinate with a field-dependent term motivated by classical ion-evaporation forms \cite{iribarne1976evaporation, loscertales1995experiments}, the fitted models provide an efficient surrogate for MD that can be evaluated locally along particle trajectories. The fitted parameters reported for each solvation number and polarity compactly encode the observed shift from thermally controlled dissociation to strong field breakup, and are designed for direct use in plume solvers that explicitly propagate metastable clusters through nonuniform acceleration fields \cite{petro_2022, smith2024propagating}.

The microcanonical ensemble trajectories resolve field-driven changes in fragmentation topology. In weak fields, trimer dissociation predominantly proceeds through neutral ion-pair evaporation to charged dimers, whereas strong fields promote direct ejection of the charged core ion, yielding monomers accompanied by double neutral clusters. Tetramers denote three major branching pathways that depend on polarity and field magnitude: single neutral evaporation, double neutral emission, and charged core ion ejection. These pathways are consistent with the idea that strong fields can align and stretch charge-separated configurations, increasing access to charge ejection channels that are suppressed at low fields \cite{kapranov2013stochastic, su2023fragmentation}. The pathway-resolved results also emphasize that higher solvation clusters cannot be represented solely by sequential single neutral pair evaporation in acceleration environments \cite{prince2015molecular}.

The pathway-resolved kinetics described in this work have immediate consequences for electrospray diagnostics and performance inference. Current-based measurements such as TOF and RPA sample only charged products, while the branching quantified here predicts significant redistribution of mass into neutral clusters during acceleration \cite{schroeder2023inferring, lyne2024inferring}. Incorporating both lifetimes and kinetic dissociation pathways into forward models can improve systematic bias that arises when fragmentation is modeled with incomplete product families or field-free rate constants. From this standpoint, the present dataset is complementary to emerging acceleration region constraints from energy-resolved measurement and multidimensional TOF \cite{lyne2023simple, ulibarri2025direct}. 

Underlying uncertainties due to the limitations in the modeling approach selected present clear extensions for future work. Fixed-charge force fields can misestimate dissociation energetics and field response due to missing electronic polarization. Polarizable force fields, charge-scaling strategies, or ab initio methods would refine absolute barriers and strengthen transfer across propellant chemistry \cite{bedrov2019molecular, doherty2017revisiting}. Additional MD campaigns at lower internal energies and larger solvation numbers would better capture cooled downstream populations and the higher mass tails observed experimentally. These developments would further solidify the connection between atomistic fragmentation dynamics and plume metrics observed experimentally to characterize electrospray sources. 

\section*{Acknowledgments}

The authors would like to graciously acknowledge the support of the Air Force Office of Scientific Research Young Investigator Program under Grant No. FA9550-23-1-0141. 

Adler Smith would like to thank the National Science Foundation Graduate Research Fellowship under Grant No. DGE–2139899 for supporting this work. 

\section*{Author Contributions}

\textbf{Nicholas Laws}: Conceptualization (lead); Data curation (lead); Formal analysis (lead); Investigation (lead); Methodology (lead); Software (lead); Validation (lead); Visualization (lead); Writing – original draft (lead); Writing – review \& editing (lead).

\textbf{Adler Smith}: Conceptualization (equal); Data curation (equal); Formal analysis (equal); Investigation (equal); Methodology (equal); Software (equal); Validation (equal); Visualization (equal); Writing – original draft (lead); Writing – review \& editing (lead).

\textbf{Elaine Petro}: Project administration (lead); Resources
(lead); Supervision (lead); Writing – review \& editing (supporting).


\nocite{*}
\bibliography{ref}

\end{document}